\documentclass[12pt]{iopart}
\usepackage{iopams,graphicx}

\newcommand{\EG}{{\textrm{e.g.}}}
\newcommand{\IE}{{\textrm{i.e.}}}
\newcommand{\EA}{{\textit{et al.}}}

\begin{document}

\title{Extracting the neutron-neutron scattering length ---
recent developments}

\author{Anders G{\aa}rdestig}

\address{Department of Physics and Astronomy,
University of South Carolina,
712 Main Street,
Columbia, South Carolina 29208,
U.S.A.
\footnote{current address: Physics Department, Whitworth University,
300 W. Hawthorne Rd., Spokane, WA 99251\\
E-mail: {agardestig@whitworth.edu}}
}
\ead{anders@physics.sc.edu}

\begin{abstract}
The experimental and theoretical issues and challenges for extracting the
neutron-neutron scattering length are discussed.
Particular emphasis is placed on recent results and their impact on the field.
Comments are made regarding current experimental and theoretical possibilities.
\end{abstract}


\submitto{\JPG}
\maketitle

\section{Introduction}

It is well known that the lightest two quarks (up and down) are almost
identical in mass, at least when compared to the scale of the strong
interaction in nuclear systems ($\sim1$~GeV).
This means that the two lightest quarks have almost identical strong
interactions, i.e., they are almost identical degrees of freedom.
While the reason for this similarity is not yet understood, the corresponding
approximate symmetry, isospin, is a remarkably effective tool in organizing and
making sense of the vast landscape of nuclei and particles.
This was realized very early on by Heisenberg, who introduced the
concept~\cite{isospin}.
Thus, it is common, and often very useful, to consider the proton and neutron
to have the same properties under the strong interaction.
This equivalence is called charge symmetry (CS), which is a special case of
isospin invariance.
Formally, CS is the invariance of the Lagrangian under a $180^\circ$ rotation
around the 2-axis in isospin space.
Its violation, fundamentally due to the different masses~\cite{wei77,mumd}
and electromagnetic
properties of the light quarks, is called charge symmetry breaking (CSB).

The effects of CSB can be detected in the properties of the hadrons.
The most obvious and most important case is the neutron-proton mass difference,
which would have opposite (i.e., negative) sign (the proton heavier, because
of the Coulomb interaction between the quarks) if it was not for the quark
mass difference.
A heavier proton would lead to a different chart of nuclides and nuclear
abundances (since Big-Bang nucleosynthesis is crucially dependent on the
relative abundances of neutrons and protons)
that would yield an entirely different universe than the one we live in.
Similar mass differences occur between other hadron isospin partners as well.
Some other important CSB signals include the small difference between the
neutron and proton analyzing power in $np$ elastic
scattering~\cite{Abegg:1986ei},
the $\rho$-$\omega$ mixing~\cite{Barkov:ac}, the binding-energy difference
between mirror nuclei (the Nolen-Schiffer anomaly~\cite{Nolen:ms}),
and the recently measured non-vanishing forward-backward asymmetry for
$np\to d\pi^0$~\cite{Allena} and the $5\sigma$ signal for
$dd\to\alpha\pi^0$~\cite{IUCFCSB}.
Also a recent analysis of $\pi N$ scattering data using chiral perturbation
theory found a small CSB effect~\cite{ulfIV}.
(Detailed reviews of this topic can be found in Refs.~\cite{MNS,MvO,MOS}.)
However, for reasons that will soon become clear, the second-most important
CSB effect (after the neutron-proton mass difference) is the inequality
of the $pp$ and $nn$ scattering lengths.

Already early on it was discovered~\cite{Bethe} that the wave function of the
two-nucleon system could be well parameterized by a few parameters,
independently of the shape of the assumed potential.
The expression commonly seen in this context is the momentum/energy expansion
of the ($S$-wave) phase shift $\delta_0$:
\begin{equation}
  p \cot\delta_0 = -\frac{1}{a}+\frac12 r_0 p^2 + O(p^4),
\end{equation}
where $p$ is the relative momentum, $a$ the scattering length and $r_0$ the
effective range.
This is the so-called effective range expansion (ERE), which has been widely
used in treatments and applications of nucleon-nucleon interactions at low
energies.

As already mentioned, CSB causes the scattering lengths of $pp$ and $nn$
systems to differ substantially.
Phenomenological calculations relate the difference between the scattering
lengths to CSB meson exchanges, e.g., mixed $\rho$--$\omega$ exchange.
An evaluation of these calculations can be found in the CSB review by Miller
and van Oers~\cite{MvO}.
The currently accepted values~\cite{NNreview} are
\begin{eqnarray}
  a^{\rm str}_{pp} & = & -17.3\pm0.4\ {\rm fm}, \label{eq:app} \\
  a^{\rm str}_{nn} & = & -18.9\pm0.4\ {\rm fm}, \label{eq:ann}
\end{eqnarray}
where the superscript `str' indicates that electromagnetic effects (mainly
Coulomb for $pp$ and magnetic-moment interaction for $nn$) have been
theoretically removed to reveal the strong interaction contribution to the
scattering lengths.
This large relative difference between the scattering lengths is a consequence
of the reciprocal relationship between the scattering lengths and the
potentials:
\begin{equation}
  \frac{1}{a_{pp}}-\frac{1}{a_{nn}} = M \int_0^\infty
  dr u_{pp}(V_{pp}-V_{nn})u_{nn},
\end{equation}
where the $u_x$ and $V_x$ are the wave functions and potentials for the
indicated nucleon pairs, with the normalization $u_x(0)=0$ and
$u_x\to1-r/a_x$ as $r\to\infty$.
It can be shown~\cite{MvO} that the scattering length difference $\Delta a$
is related to the CSB potential via an enhancing factor of 10--15:
\begin{equation}
  \frac{\Delta a}{a} = \frac{a_{pp}-a_{nn}}{a} =
  (\mbox{10--15})\frac{\Delta V_{\rm CSB}}{V}.
\end{equation}
This enhancement is because of the large value of the scattering lengths
(compared to the range of the strong interaction)---an example of fine tuning.
Thus, the scattering length difference is very sensitive to the finer details
of the $NN$ interaction, a fact that is used to pin down the CSB part of the
modern high-precision $NN$ potentials~\cite{NijmPot,CDBonn,AV18}.
These potentials are in turn used in very precise calculations of the energy
levels of low-mass ($A<14$, say) nuclei~\cite{lowAcalc,NCSM}.
The corresponding calculation of the tritium--helium-3 binding-energy
difference is connected to the sign of $\Delta a$, a fact which has long been
known~\cite{bind3N}.
Thus, with the currently accepted values for $a^{\rm str}_{pp}$ and
$a^{\rm str}_{nn}$ [Eqs.~(\ref{eq:app}) and (\ref{eq:ann})], the theoretical
predictions are in very good agreement with experiments~\cite{lowAcalc}.
The charge dependent contribution to the $^3{\rm H}$-$^3{\rm He}$ binding
energy difference is calculated to be 64~keV, to be compared to the total
difference 757~keV.
This is in good agreement with the experimental value 764~keV.
In order to achieve this it is necessary to include state-of-the-art
three-nucleon forces (3NF), such as the Urbana IX~\cite{UIX},
Illinois~\cite{IL}, Brazil~\cite{Brazil}, Texas~\cite{Texas} and
Tucson-Melbourne (TM)~\cite{TM3NF}.
However, if the sign of $\Delta a$ was reversed, \IE, $a_{pp}$ more negative
than $a_{nn}$, this agreement would be off by more than 100~keV, since
the CSB part of the potential would change sign.
See also the later calculations by Nogga \EA~\cite{Nogga3NF} showing excellent
agreement between results obtained using Faddeev equations with
AV18~\cite{AV18}+Urbana IX~\cite{UIX}
and using hyperspherical harmonics with CD-Bonn~\cite{CDBonn}+TM~\cite{TM3NF}.

There are two types of difficulties that have to be overcome to extract the
proper scattering lengths.
First, the electromagnetic interactions have to be removed, which can only be
done theoretically, using sophisticated state-of-the-art tools.
These corrections are huge for the proton-proton case---the uncorrected $pp$
scattering length is -7.8063~fm~\cite{app}.
The calculation of the corrections is far from trivial, although it is
well-known how to proceed.
These extractions introduce a small model-dependence through the assumptions
made about form factors~\cite{app}, i.e., off-shell behavior.
The electromagnetic corrections for $a_{nn}$ are naturally much smaller and are
dominated by the magnetic moment - magnetic moment interaction.
Calculations~\cite{MNS} reveal that this correction is -0.3~fm.
Second, the unavailability of dense enough free-neutron targets makes direct
measurements almost impossible.
Nevertheless, there have been attempts to use direct $nn$ scattering to
determine $a_{nn}$, most recently one under way in Snezhinsk~\cite{yaguar}.
These ideas, which so far have not yielded any results, will be discussed in
more detail in Sec.~\ref{sec:direct}.
Because of this difficulty, our present knowledge of the $nn$ scattering length
is based exclusively on information gleaned from indirect reactions.
The two main processes that have been used are $nd\to nnp$ and
$\pi^-d\to nn\gamma$, both of which have a long and interesting history.
They will be discussed at length in Secs.~\ref{sec:nd} and \ref{sec:pid},
respectively.

This topical review will cover the efforts to extract $a_{nn}$ over the last
few decades, with particular focus on the developments since the
review by Slaus, Akaishi, and Tanaka~\cite{Slausetal}.
Some of the later results have been discussed in
Refs.~\cite{NNreview,Howellreview}, but no up-to-date comprehensive review is
currently available.
It is the purpose of the present work to fill in this gap in the literature
and bring the topic up to date.
In an upcoming, lengthier, review also the earlier results and other
specific issues (like CSB and chiral perturbation theory) will be covered in
more depth~\cite{AGcompreview}.

\section{Indirect methods}
The main idea behind the indirect measurements of $a_{nn}$ is to find
a nuclear reaction which produces two free neutrons in a final state with
low relative energy.
By detecting neutrons in such a configuration [the final state interaction
(FSI) region where the scattering length plays an important role], it is
possible to extract a value for the scattering length.
This is however only possible if a) the detection of the neutrons, including
full kinematic information, is possible to sufficient precision and efficiency,
and b) the corresponding theoretical calculation has small theoretical error
bars.
Regardless of the reaction chosen, the extraction depends on theory and
particularly on the sensitivity to $a_{nn}$ in the chosen theoretical
framework.
Since neutron detection is notoriously inefficient, it is often advantageous
to detect instead another, more easily detected particle, such as a proton or
photon, to constrain the relative energy and momentum of the neutrons.
Ideally, this third particle should have a comparatively weak interaction with
the neutrons, so that its presence has limited influence on the process.
This has led to proposals and experiments using electroweak processes,
in particular the $\pi^-d\to nn\gamma$ reaction.
However, the bulk of experiments have relied on the basic (though far
from trivial) deuteron breakup reaction $nd\to nnp$.
This process suffers from the possible complications of three-nucleon forces,
which might be part of the reason for the conflicting results obtained for
$nd$ and $\pi^-d$ experiments~\cite{Slausetal}.
The deuteron breakup reaction has, throughout its entire (including recent)
history, been suffering from significant disagreements between experiments.
For this reason we treat this reaction first and temporarily postpone the
discussion of the electroweak probe experiments.

\subsection{Extractions using $nd\to nnp$}
\label{sec:nd}
The deuteron break-up reaction $nd\to nnp$ has a long, unfortunate history of
conflicting results.
For a more detailed description of the earlier results see the previous
reviews~\cite{Slausetal,NNreview,Howellreview},
or the upcoming longer version of the present work~\cite{AGcompreview}.
The difficulties are both experimental and theoretical.
Huhn \EA,~\cite{Huhn1,Huhn2} point out that the kinematically incomplete
measurements in general extract $a_{nn}\sim-19$~fm, while the kinematically
complete ones end up in the neighborhood of $-16.5$~fm.
However, a reanalysis of the kinematically incomplete experiments using
state-of-the-art three-nucleon formalisms, including modern high-precision
$NN$ potentials and three-nucleon forces [of the two-meson exchange
Tucson-Melbourne (TM)
kind~\cite{TM3NF}] results in values around $-15.5$~fm~\cite{Tornowetal}.
This drastic change in the kinematically incomplete experiments points to a
strong dependence on the $NN$ force used in the calculations.
In the complete kinematics results, however, the shape of the FSI peak is
essentially determined by the value of $a_{nn}$  with much reduced dependence
on the choice of the $NN$ potential~\cite{Huhn2}.

The influence of $NN$ and three-nucleon forces (3NF) was studied by
Wita{\l}a \EA~\cite{Witala}, who found that for certain configurations the
FSI peak is practically insensitive to the choice of $NN$ interaction.
Also the effects of 3NF could be minimized at these angles.
Note, however, that this study used the TM 3NF only.

With the modern development of effective field theories (EFTs) for the strong
interaction of nucleons (initiated by Weinberg~\cite{Weinberg}), it is now
possible to include the three(and four)-nucleon forces in the same theoretical
framework as the two-nucleon force.
This is not possible in the phenomenological approach mentioned above.
One of the most important aspects here is that an EFT allows for an expansion
in a small parameter.
For the cases of interest for this review, the EFT of choice is heavy baryon
chiral perturbation theory (HB$\chi$PT), with an expansion parameter
$\chi=Q/\Lambda_\chi$.
Here $Q\sim m_\pi$ is a typical energy or momentum of reactions involving
low-energy nucleons and pions and
$\Lambda_\chi\sim m_N\sim 4\pi f_\pi\sim 1$~GeV is the scale at which the
(chiral) expansion breaks down.
If the relevant energies $Q$ are sufficiently smaller than $\Lambda_\chi$
the expansion should converge.
In addition, $\chi$PT organizes the different contribution to CSB in a
hierarchal structure, with the lowest order CSB contribution to the nuclear
force being determined by the $a_{nn}-a_{pp}$ difference~\cite{vKFG}, again
emphasizing the fundamental importance of this parameter.
See, e.g., Ref.~\cite{ulfreview} for a thorough introduction to and further
details about HB$\chi$PT.
Significant progress has also been made for few-nucleon
systems~\cite{Evgeny2N3N}, but no one has yet done chiral calculations
needed for the extraction of $a_{nn}$ from $nd$ breakup.
Given the achievements of EFT calculations in general (see, e.g.,
Ref.~\cite{DanielJPG} for the use of EFT in processes involving
electromagnetic probes) and for $a_{nn}$ extraction via the two-nucleon
reactions in particular (see Secs.~\ref{sec:pid} and \ref{sec:otherindirect})
the natural next step is to use EFT-based calculations also in extracting
$a_{nn}$ from observables in three-nucleon systems.

\begin{figure}[P]
\includegraphics*[width=6in]{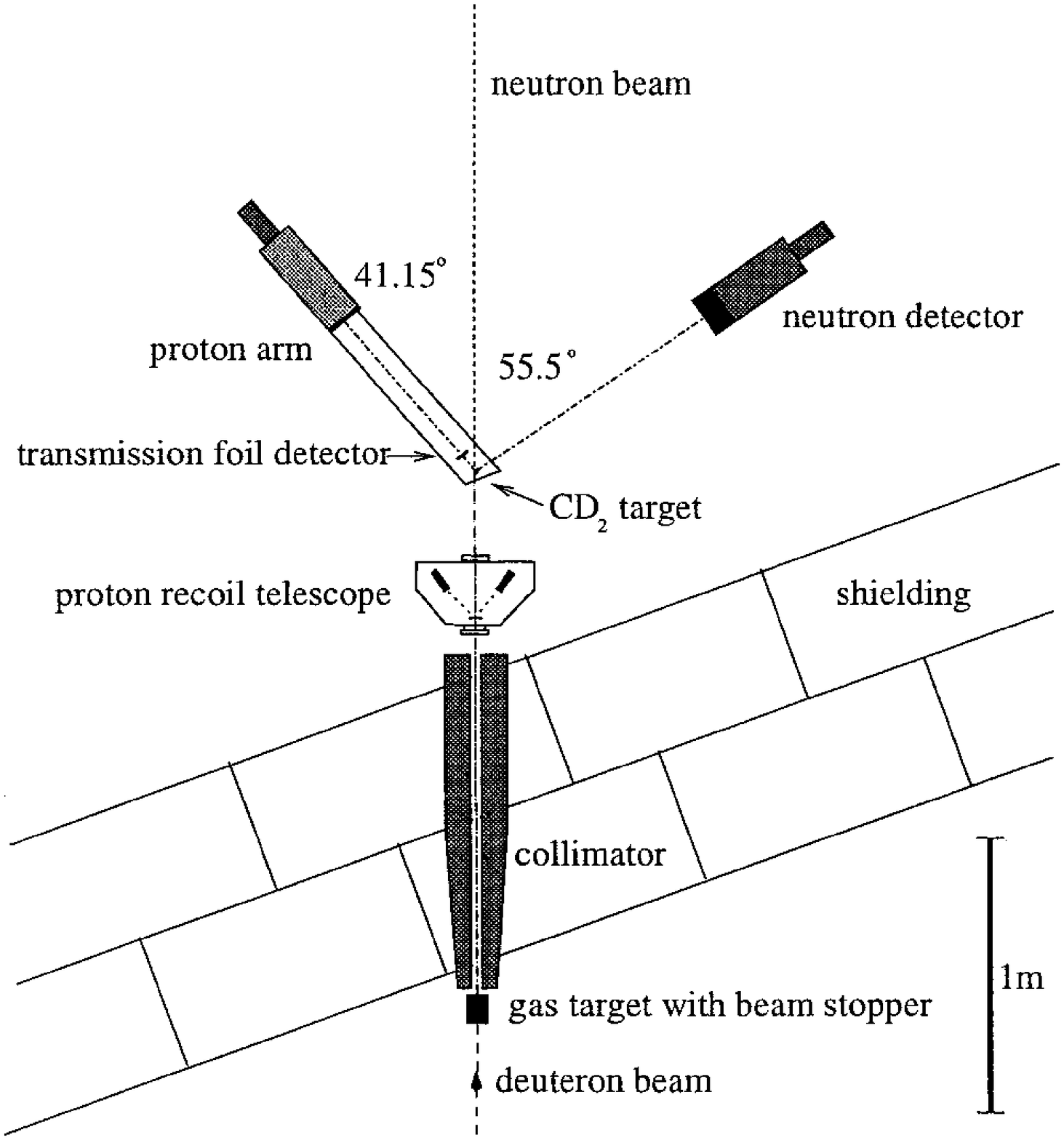}
\caption{The set-up for the Bonn $nd$ experiment.
Reprinted with permission from V. Huhn \EA, Phys.\ Rev.\ C {\bf 63} (2000) 014003. Copyright (2000) by the American Physical Society.}
\label{fig:Bonn}
\end{figure}

The most recent experimental results come from two different groups.
One group, based in Bonn, measured the $d(n,pn)n$ reaction at neutron beam
energies of 16.6 and 25.3~MeV~\cite{Huhn1,Huhn2} using the set-up shown in
Fig.~\ref{fig:Bonn}.
They claimed a practically model-independent extraction of $a_{nn}$.
Their result obtained from the absolute cross section in the FSI peak region is
$-16.3\pm0.4$~fm (16.6~MeV).
They also reported results from the relative cross section at the FSI peak
(normalized in the $np$ FSI region):
$-16.1\pm0.4$ (25.3~MeV) and $-16.2\pm0.3$~fm (16.6~MeV).
In order to facilitate comparisons they extracted also the $np$ scattering
length at 25.2~MeV in good agreement with other determinations.
Note however, that the $a_{np}$ experiment is performed with the proton
detector replaced with another neutron one.
Hence, this cross check is strictly speaking not done in exactly the same
experimental set-up as for $a_{nn}$.

The theoretical calculations were performed using three-body Faddeev
equations with the CD-Bonn $NN$ potential~\cite{CDBonn}.
Inclusion of the TM 3NF caused no significant changes.

\begin{figure}[P]
\includegraphics*[width=6in]{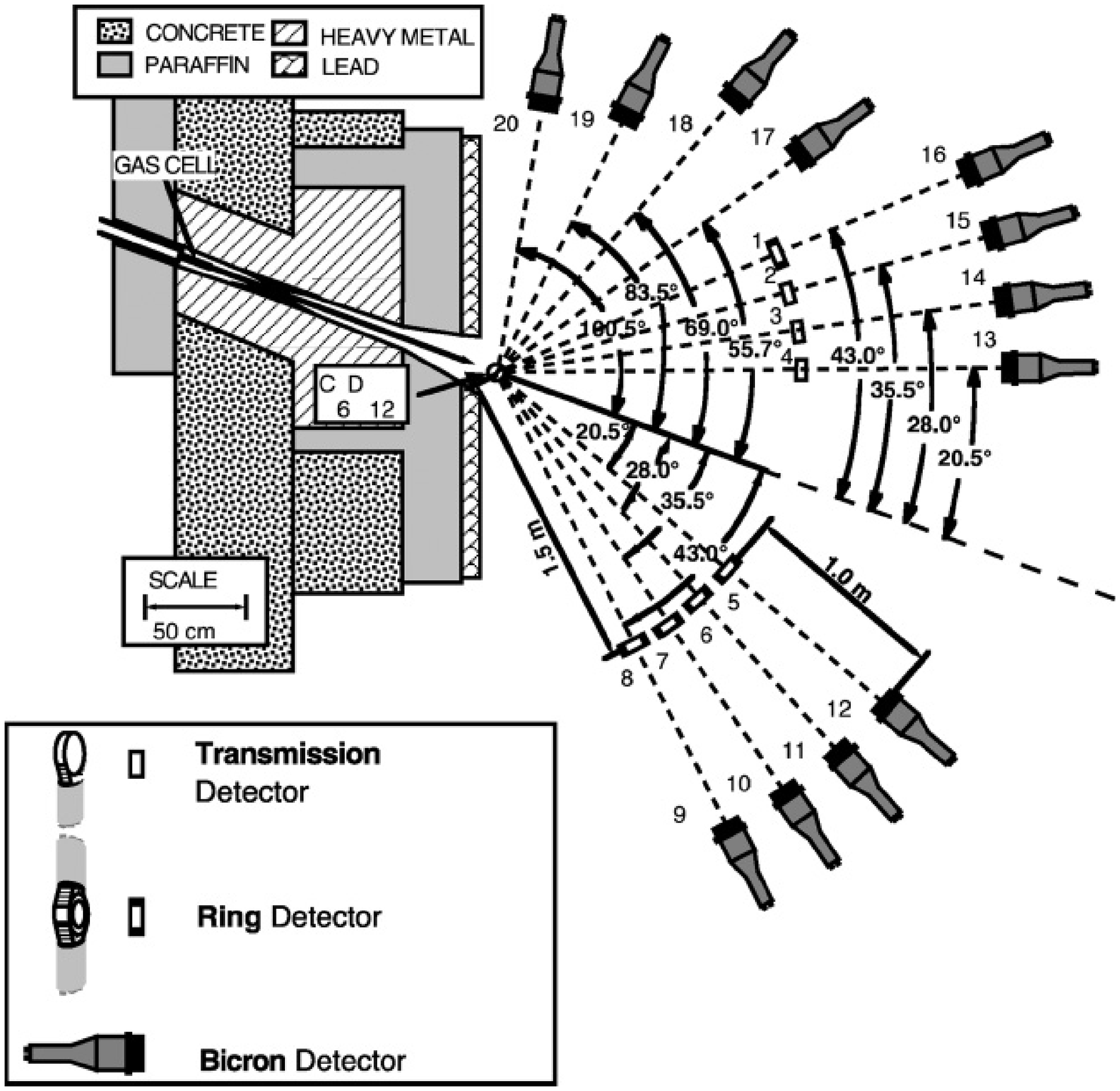}
\caption{The set-up for the TUNL $nd$ experiment.
Reprinted with permission from D. E. Gonzalez \EA, Phys.\ Rev.\ C {\bf 73} (2006) 034001. Copyright (2006) by the American Physical Society.}
\label{fig:TUNL}
\end{figure}

At about the same time, a TUNL group measured the $d(n,nnp)$ reaction, i.e.,
an over-complete experiment with all final particles
detected~\cite{TUNL1,TUNL2}.
This set-up is shown in Fig.~\ref{fig:TUNL}.
The incident neutron energy here was 13~MeV.
Using the same theoretical framework as the Bonn group,
they arrived at $a_{nn}=-18.7\pm0.6$~fm and $a_{np}=-23.5\pm0.8$~fm.
The extraction of $a_{np}$ in this case was done using the same detectors as
used for extracting $a_{nn}$.
It had been shown earlier, using the TM 3NF~\cite{TM3NF}, that the relative
changes due to the 3NF were identical in the $nn$ and $np$
peaks~\cite{Coonetal}.
The well-known value for $a_{np}$ could hence be used to put limits on the
influence of the 3NF in the $nn$ FSI region.

The experimental results were compared to three-body Faddeev calculations
using the modern $NN$ potentials: AV18~\cite{AV18}, CD-Bonn~\cite{CDBonn},
Nijm I and Nijm II~\cite{NijmPot}.
The three-nucleon force effects were estimated using the TM 3NF.

A second simplified experiment was carried out very recently at Bonn.
This time only the final proton was measured giving
$a_{nn}=-16.5\pm0.9$~fm~\cite{vonW}.
Once again, the calculations were done with the same methods as before.

The original Bonn result for $d(n,np)n$ $a_{nn}=-16.1\pm0.4$~fm~\cite{Huhn2}
together with its follow-up, simplified version
$d(n,p)nn$ $a_{nn}=-16.5\pm0.9$~fm~\cite{vonW} thus differ with more than
3$\sigma$ from the TUNL result $d(n,nnp)$: $-18.7\pm0.7$~fm~\cite{TUNL1,TUNL2}.

The main experimental difference between the two groups' investigations is the
chosen geometry for the $nn$ system: Bonn detected one neutron (and the proton)
in what they called recoil geometry, i.e., the neutron and proton on separate
sides of the beam.
TUNL, on the other hand, used final-state geometry, where both neutrons are
detected close together on the same side of the beam.
Also, while TUNL extracted $a_{nn}$ and $a_{np}$ simultaneously, the Bonn
set-up needed to change detectors to make the switch to the measurement of $a_{np}$.

The different geometries could also be a cause of theoretical concern.
There is a (perhaps small) possibility that the predictions of the final-state
geometry cross section is wrong~\cite{Huhn2}.
One should also make certain that other implementations of 3NF (Urbana,
Illinois, Texas, Brazil, and chiral) behave in a similar way.
To my knowledge this has not been done.

The Bonn experiment is currently being set-up at TUNL so that both experiments
can be run at the same time, using the same neutron beam~\cite{Howellperscom}.
This is being done in order to reduce the possibility of experimental errors.

\subsection{Extractions using $\pi^-d\to nn\gamma$}
\label{sec:pid}
The $\pi^-d\to nn\gamma$ reaction has a long history, see
Refs.~\cite{Slausetal,AGcompreview} for details.
In this case a slow pion is captured in an atomic orbital around the deuteron
and cascades down until it reaches an $s$ orbital.
Its wave function then has a finite overlap with the deuteron allowing for the
pion absorption and the subsequent breakup into two neutrons and a photon to
occur.
This process has the advantage that the third final state particle is an
easily measured high-energy ($\sim130$~MeV) photon, which has weak final
state interaction with the two neutrons.
Since the shape of the spectrum is less sensitive to systematic errors than
the absolute capture rate, the extraction is in this case always done
by fitting the shape of theoretical calculations, with varying $a_{nn}$,
to the data.
The fact that the pion is captured at rest in an atomic orbital around the
deuteron has the further theoretical advantage that only the first ($F_1$) of
the Chew-Goldberger-Low-Nambu (CGLN) amplitudes~\cite{CGLN} can contribute,
since the pion momentum is negligible on the nuclear scale.
This amplitude corresponds to zero-momentum pions interacting with the proton
(or $\gamma n\to p\pi^-$ at threshold).

\subsubsection{Earlier theoretical methods}
In order to obtain $a_{nn}$ from $\pi^-d\to nn\gamma$, two different
theoretical approaches have been used, both employing single-nucleon dynamics
(impulse approximation).
Neither attempted a full treatment of two-nucleon effects, though some steps
were taken in both approaches to at least estimate their influence.

The first (phenomenological) method is exemplified and dominated by the
thorough calculations of Gibbs, Gibson, and Stephenson (GGS)~\cite{GGS}.
In their results, the one-body amplitude (the Kroll-Ruderman term) is
calculated with first order relativistic corrections.
The neutron-neutron wave function was obtained by starting from the asymptotic
state and then integrating in from large $r$ using the Reid soft-core
potential~\cite{Reid}.
Inside of $r=1.4$~fm a fifth degree polynomial with appropriate matching and
boundary conditions was used.
The theoretical error was estimated after an extensive study of the influence
of many sources.
For small relative momenta ($<48$~MeV or an $nn$ opening angle $<30^\circ$)
the error is dominated by the uncertainties in the short-range part of the
neutron-neutron wave function.
This was estimated by changing the shape of the $nn$ wave function below 1.4~fm
and resulted in an error of $0.3$~fm in $a_{nn}$.
This small error depends crucially on the small relative energy, \IE,
the applicability of these calculations is restricted to the FSI region only.

In the second calculation method the neutron FSI was analyzed in terms of
Muskhelishvili-Omn\`es dispersion relations.
This work was done by de T\'eramond, Gabioud and
collaborators~\cite{deTeramond}.
In their second paper, they included pion rescattering terms and off-shell
effects.
This allowed them to show that these subtleties have importance for large
($>80$~MeV) relative $nn$ momenta.
The final paper of de~T\'eramond and Gabioud included higher partial waves and
showed that they contributed very little to the extracted $a_{nn}$, thereby
confirming the validity of the approximations made in many of the previous
calculations.
The total theoretical error in these calculations is estimated to be
 $\Delta a_{nn}=0.2$~fm.

While these theoretical error bars are quite impressive, there are still some aspects that
need improvement.
In both cases only part of the measured spectrum (the FSI peak) is used in the analysis,
which raises some concerns regarding the fitting procedure.
By fitting only half the spectrum, we loose not only statistical precision but also
any additional information in the quasi-free peak.
Also it is somewhat questionable if
By taking advantage of modern development of effective field theory it should be possible
to do a systematic calculation, where all uncertainties can be
adequately described in a consistent manner within the same theoretical framework.
That it is indeed possible to perform a calculation that successively addresses
all these issues will be shown below in Sec.~\ref{sec:moderntheory}.

\subsubsection{Experiments}
The most recent experiments were carried out at Paul Scherrer Institute (PSI)
[then the Schweizerisches Institut f\"ur Nuklearforschung
(SIN)]~\cite{PSI, Schori} in the eighties and at the Clinton P. Anderson
Meson Physics Facility at Los Alamos (LAMPF)~\cite{LAMPF} in the nineties.
In the first PSI experiments~\cite{PSI}, only the photon was detected either
using a converter and electron-positron pair spectrometer or directly in
lead-glass crystals.
The combined need of measuring both the shape and position of the FSI peak
(both are sensitive to the value of $a_{nn}$) was overcome by
alternating with the detection of $\pi^-p\to n\gamma$, which provided a
calibration line for the $\pi^-d$ spectrum.
The final result was $-18.5\pm0.4$~fm, including both theoretical and
experimental uncertainties in the error bar.
This value is independent of the choice of theory used.
A later experiment measured also one of the neutrons in coincidence and arrived
at the somewhat less precise result $-18.7\pm0.6$~fm~\cite{Schori}, but in
good agreement with the earlier value.
This removed concerns about systematic errors, since this experiment used
time-of-flight (TOF) detection of one neutron in coincidence with a photon,
rather than detecting the photon alone.

\begin{figure}[P]
  \includegraphics*[width=6in]{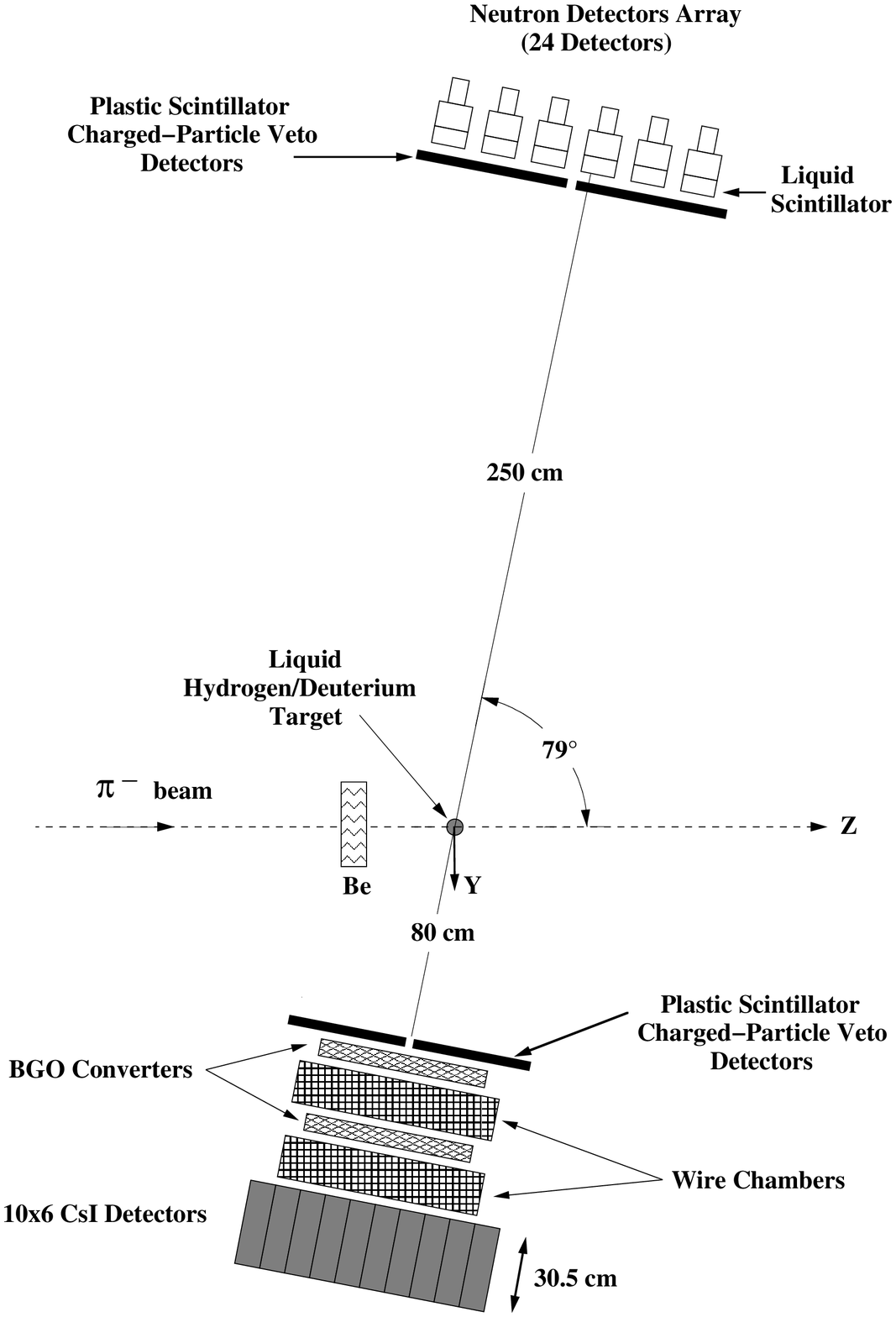}
\caption{The set-up for the LAMPF $\pi^-d\to nn\gamma$ experiment.
Reprinted with permission from Q. Chen \EA, Phys.\ Rev.\ C {\bf 77} (2008) 054002. Copyright (2008) by the American Physical Society.}
\label{fig:LAMPFsetup}
\end{figure}

\begin{figure}[P]
  \includegraphics*[width=6in]{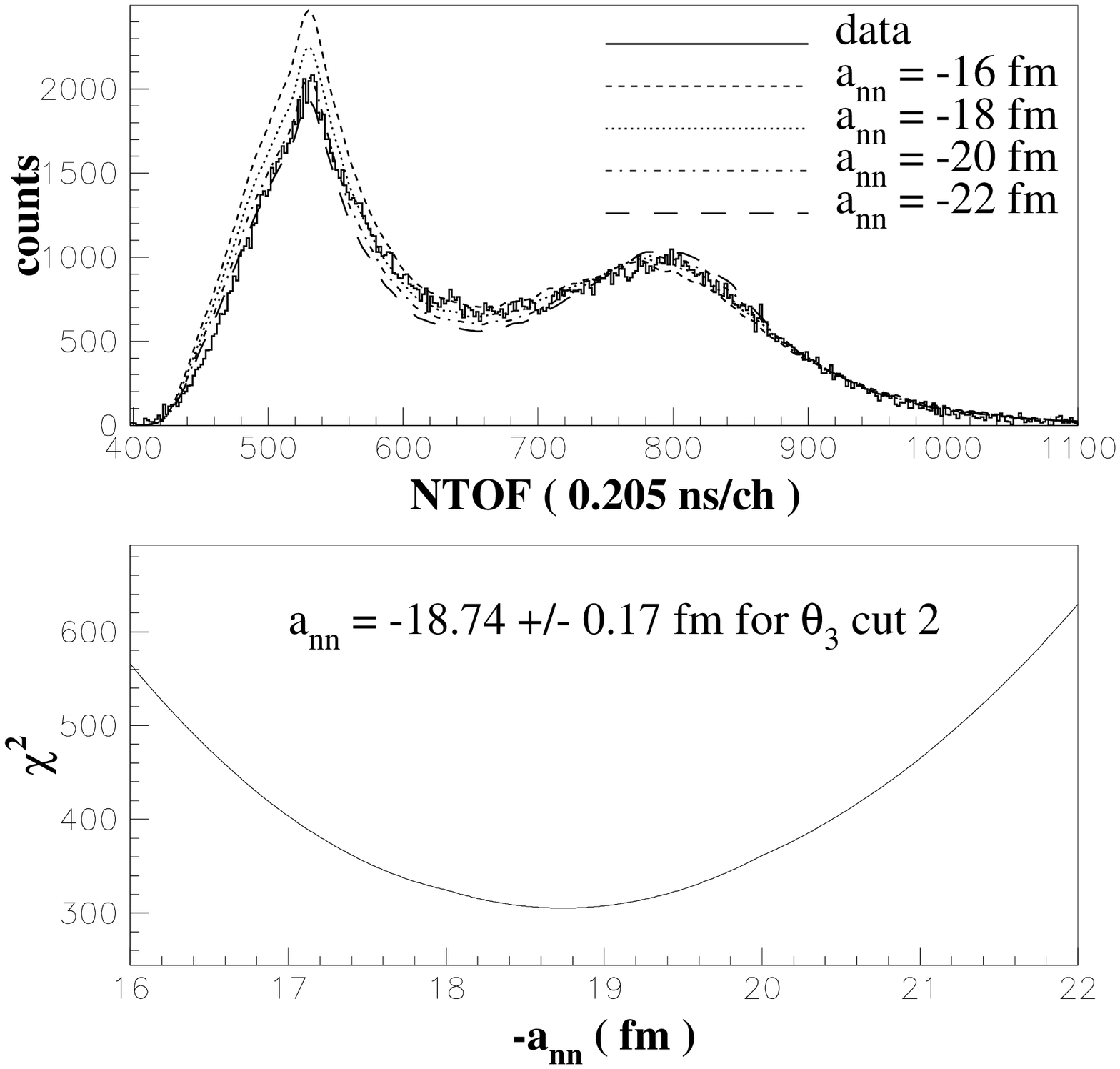}
\caption{Result for the LAMPF $\pi^-d\to nn\gamma$ experiment for the angular
range $0.05<\theta_3<0.10$ (rad).
Similar results were obtained for other angular ranges.
Reprinted with permission from Q. Chen \EA, Phys.\ Rev.\ C {\bf 77} (2008) 054002. Copyright (2008) by the American Physical Society.}
\label{fig:LAMPFresults}
\end{figure}

The LAMPF experiment followed the same principle of a precise neutron TOF
measurement.
Instead of using pair conversion, the photon was now detected by BGO
scintillators complemented by wire chambers and CsI detectors to give accurate
position information.
The layout of the experiment is given in Fig.~\ref{fig:LAMPFsetup}.
Their result was originally reported as
$a_{nn}=-18.50\pm0.05({\rm statistical})
\pm0.44 ({\rm systematic})\pm0.30({\rm theoretical})$~fm~\cite{LAMPF},
which was later modified by using relativistic phase space and a
more accurate fitting procedure yielding $a_{nn}=-18.63\pm
0.10({\rm statistical})\pm0.44 ({\rm systematic})\pm0.30({\rm theoretical})$~fm
\cite{Howell2}.
An example of their measurements is given in Fig.~\ref{fig:LAMPFresults}.

Thus, both experiments gave very similar results, regardless of theory,
resulting in the remarkably low theoretical error of 0.3~fm in $a_{nn}$
extracted from $\pi^-d\to nn\gamma$.
Given the stability of these results and the theoretical appeal of having
$|a_{nn}|>|a_{pp}|$~\cite{bind3N}, the value $-18.6\pm0.3$ has been used as
the accepted value for some time now~\cite{NNreview}.

\subsubsection{Modern theoretical treatment}
\label{sec:moderntheory}
Since the early nineties effective field theory (EFT) has proved to be a very
successful tool to describe few-nucleon systems.
Particularly useful is the extension of chiral perturbation theory ($\chi$PT)
to the nucleon sector (heavy baryon chiral perturbation theory, HB$\chi$PT)
which was pioneered by Weinberg~\cite{Weinberg} (see, \EG,
Ref.~\cite{ulfreview} for a review).
The advantage of this approach is manifold:
HB$\chi$PT has strong connections to the underlying theory of the strong
interaction (QCD); in particular it inherits the chiral symmetry of the light
quarks.
Being an EFT, it provides a hierarchal structure of the entire calculation,
a so-called power counting, which furnishes the means of successive
improvements of the calculation when needed.
The power counting also enables us to make well-defined error estimates.
The drawback of the method, but also its strength, is that at (almost) each
higher order new coupling parameters, so-called low-energy constants (LECs)
are introduced.
These have to be fixed by comparison with experiments, but since the same LECs
apply to a variety of different processes, once they are determined from one
of them, they can be used in all the other ones.

In some recent papers, the HB$\chi$PT has been successfully applied to the
$\pi^-d\to nn\gamma$~\cite{GP1,GP2,AG1}.
In addition to the photon-pion-nucleon amplitudes, also the bound state
(deuteron) and scattering state (neutron-neutron) wave functions are evaluated
in the same $\chi$PT framework.
This follows the ideas originally developed by Phillips and Cohen~\cite{PC} for
the deuteron, which were extended to cover the neutron-neutron state in
Ref.~\cite{GP1}.
The starting point of this approach are the well-known asymptotic states
given by $A_S$ and $\eta=A_D/A_S$ for the deuteron ($A_S$ and $A_D$ are the
asymptotic $S$ and $D$ state normalizations) and $a_{nn}$ and $r_{nn}$
for $nn$ scattering.
Then the nucleon-nucleon force (here of course the chiral $NN$
one- and two-pion exchange potential~\cite{TPEP}) is introduced and the
Schr\"odinger equation is used to integrate in from the asymptotic state.
At some distance $R$ in the few-fermi range, the calculation runs into physics
which is not defined in $\chi$PT.
The integration is therefore truncated at this point and matched to a spherical
well solution for the remainder ($r<R$).
This procedure provides a way to parameterize and regularize our ignorance of
short-distance physics.
Because of renormalization invariance the end result should be independent of
the choice of $R$.

The calculations originally included everything up to next-to-next-to-leading
order (NNLO), which takes the impulse approximation to the next level
and also includes the first two-body currents.
The higher order one-body terms, which includes a few unknown LECs are fixed
by separate calculations for the single-nucleon system, performed earlier in
Ref.~\cite{Fearing}.
The two-nucleon amplitudes do not introduce any new LECs at this order.
Within this approach the authors proved it possible to extract $a_{nn}$ with
a precision at the $0.2$~fm level if the fit to experimental data is restricted
to the FSI peak only.
This error in $a_{nn}$ was due to the uncertainties in the short-range part
of the $nn$ wave function, \IE, there was a small but noticeable
$R$-dependence.

Later this work was significantly improved upon~\cite{GP1,AG1}, using the
realization that at the next higher order (N3LO) there is a single counter
term, usually labeled $\hat{d}$, which renormalizes the spin-isospin flip
two-body current, i.e., the Gamow-Teller (GT) transition.
This same transition, and hence $\hat{d}$, plays an important role also in the
precise calculations of proton fusion~\cite{GTpot,Parkpp,Parkhep,BCpp}, the
$hep$ process~\cite{Parkhep}, neutrino-deuteron breakup
reactions~\cite{nud,BCpp}, tritium beta decay~\cite{GTpot,Parkhep}, muon
capture on the deuteron~\cite{apkm02,GPKM}, pion $p$-wave production on two
nucleons~\cite{HvKM}, and the corresponding part of the chiral three-nucleon
force~\cite{HvKM,SofiaLEC}.
By normalizing the $\pi^-d\to nn\gamma$ capture rate to the value
of the GT matrix element of $pp\to de^+\nu_e$ (in turn normalized by
phenomenological calculations of tritium beta decay~\cite{GTpot}), it was
possible to reduce the error in the extracted $a_{nn}$ to 0.05~fm if fitted
in the FSI region, and to 0.3~fm if the entire spectrum is fitted\footnote{An ab
initio (no-core shell model) calculation of tritium beta decay using $\chi$PT
recently appeared in the literature~\cite{SofiaLEC} and should be considered in
future chiral extractions of $a_{nn}$ from $\pi^- d\to nn\gamma$.}.
Some further details of this calculation are provided in Ref.~\cite{AG1}.

We are now completing a calculation~\cite{GPKM} that aims to investigate the
influence of recoil corrections and the consistency between higher order
corrections in the amplitudes and the wave functions.

Hence it is now possible to extract $a_{nn}$ with an error which is dominated
by the experimental uncertainties.
In addition, both theoretical and total errors would then be smaller than
the corresponding ones for the proton-proton scattering length.

\subsection{Other indirect methods}
\label{sec:otherindirect}
Here I will give a brief overview of other possible candidates to extract
$a_{nn}$ using final state interaction.

\subsubsection{$\mu^-d\to nn\nu_\mu$}
The reaction $\mu^-d\to nn\nu_\mu$ would be an ideal candidate to determine
$a_{nn}$: there are only two strongly interacting particles in the final state.
However, as pointed out in Ref.~\cite{Slausetal}, only a small fraction of the
total phase space is sensitive to $a_{nn}$, cf.\ the very similar kinematics of
the $\pi^-d\to nn\gamma$ reaction.
This means that a measurement of the total capture rate is not sufficient
to determine the scattering length; the $nn$ energy distribution also needs to
be measured.
Given the small cross section of neutrino detection, the prospect of using
this reaction to extract $a_{nn}$ seems rather slim at the present time.
However, measuring the muon capture on the deuteron has intrinsic value.
It is the simplest weakly interacting nuclear system that can both be measured
and calculated to high precision.
It has also been proposed as a tool to extract the pseudo-scalar form factor
$g_P$.

An additional characteristic of this reaction which has more immediate interest
is that it goes largely via a Gamow-Teller transition.
As already mentioned, the two-body current, as defined in the EFT framework
outlined above, depends on the same LEC $\hat{d}$ as does the calculation of
$\pi^-d\to nn\gamma$~\cite{apkm02}.
It might hence be possible to extract $\hat{d}$ directly
from a measurable and calculable two-nucleon process: $\mu^-d\to nn\nu_\mu$.
As mentioned, $\hat{d}$ renormalizes not only
pion radiative capture, but also $pp$ fusion~\cite{GTpot,Parkpp,Parkhep,BCpp},
neutrino-deuteron breakup~\cite{nud,BCpp},
tritium beta decay~\cite{GTpot,Parkhep} the $hep$ reaction~\cite{Parkhep}, as
well as $p$-wave pion production on two nucleons~\cite{HvKM} and
part of the chiral three-nucleon force~\cite{HvKM,SofiaLEC}.
This would put also these other calculations on a more solid footing.
A precise measurement of muon capture would hence indirectly help the
extraction of $a_{nn}$ from $\pi^-d\to nn\gamma$.

This realization provides part of the motivation for the current effort
to measure $\mu^-d\to nn\nu_\mu$ to 1.5\% (or better) at PSI~\cite{Kammel}.
This experiment is the latest effort of the very successful muon capture
program at PSI~\cite{muoncapturePSI}.
It has also motivated a renewed study of the EFT calculation of
$\mu^-d\to nn\nu_\mu$~\cite{GPKM}.
In this revised calculation (an extension Ref.~\cite{apkm02}), also the
deuteron and $nn$ wave functions are calculated within the EFT framework,
using the same integrating-in procedure used in the modern
chiral calculation for $\pi^-d\to nn\gamma$.

\subsubsection{$\gamma d\to nn\pi^+$}
There have been some suggestions in the literature that the
$\gamma d\to nn\pi^+$ could be used to extract $a_{nn}$.
A superficial inspection seems to suggest that the calculation involved would be very similar to that for the related (by crossing symmetry) reaction $\pi^-d\to nn\gamma$ we just discussed.
However, in this case the pion momentum is non-vanishing and all four
CGLN amplitudes contribute.
Another complication is that, with a pion in flight, constraints on the
two-nucleon wave functions (due to the Pauli principle) are different from the
atomic pion absorption case and nucleon rescattering becomes
important~\cite{Le05}.
The fact that there are now three strongly interacting particles in the final
state, raises concerns about the treatment of pion rescattering.
However, with slow enough pions, $\chi$PT tells us that this should be
a small effect that is well understood and under control.
Reference~\cite{Le05} claims a final theoretical error of 0.10~fm in $a_{nn}$
extracted using this method.
This is accomplished through a specific experimental separation of
the quasi-free and FSI regions, possible only in carefully chosen angular
configurations.
They also pointed out the importance of recoil corrections, which might also
play a role in $\pi^-d\to nn\gamma$.

This reaction could be measured at the tagged-photon set-up at MAXLab in Lund,
Sweden or at the HI$\gamma$S facility at TUNL once appropriate mirrors have
been developed~\cite{Howellperscom}.

\subsubsection{$^2{\rm H}(d,{}^2{\rm He}^2n)$}
Recently a $^2{\rm H}(d,{}^2{\rm He}^2n)$ experiment was carried out at
KVI in Groningen~\cite{Baumeretal}.
The two initial deuterons are transformed, by a Gamow-Teller transition
(spin and isospin flip) into a proton pair (unbound $^2{\rm He}$) and a
neutron pair ($^2n$).
By detecting the two protons with low relative energy ($\epsilon<1$~MeV),
one makes certain that they are in a $^1S_0$ state, i.e., $^2{\rm He}$.
Then the $nn$ pair will also be predominantly in the $^1S_0$ state.
In this particular configuration, the measurement of the protons yields a
kinematically incomplete description of the $nn$ system, which can be used to
learn more about $a_{nn}$.
After calculating the reaction using the impulse approximation,
Ref.~\cite{Baumeretal} arrived at an upper bound $a_{nn}<-18.3$~fm (95\% CL).
This experiment is hence in agreement with the accepted value and the TUNL
results for $nd$ breakup but in disagreement with the Bonn result.
Further work, especially a more sophisticated calculation using three- and
four-nucleon dynamics, is necessary for a precise extraction of $a_{nn}$ using
this method.

\section{Direct methods}
\label{sec:direct}
Despite the unavailability of dense free-neutron targets, there have been
a few suggestions to use direct $nn$ scattering.
An early suggestion was to use two simultaneous underground nuclear
explosions~\cite{Moravcsik,Glasgow}.
The intense nucleon flux would be collimated into two colliding beam and the
scattering cross section measured.
This would give a precision of about 3\% or 0.5 fm in $a_{nn}$.
While a preliminary study was carried out in 1986~\cite{Slausetal}, there are
currently no plans to pursue this approach any further.

It has also been proposed (a long time ago) to launch a pulsed nuclear reactor
into orbit, with the purpose of eliminating neutron scattering against
atmospheric nuclei~\cite{Bondarenko}.
However, the estimated uncertainty in $a_{nn}$ is about 10\% or
$2$~fm~\cite{Slausetal} and this suggestion has not been followed up.

Another possibility would be to use an intense neutron flux from
Earth-bound reactors~\cite{nnbeam}.
A recent variation of this idea is pursued by the DIANNA collaboration
using the pulsed reactor YAGUAR in the former nuclear-weapon city Snezhinsk
in Russia~\cite{yaguar}.
By triggering fusion in a cylindrical reactor containing a uranium salt
dissolved in water, the emitted neutrons get moderated by plastic walls as
they reach the hollow center of the reactor.
Detecting the resulting neutron spectrum, one can determine the $nn$ cross
section and thus the scattering length.
The advantages of this set-up are that the neutrino flux is intense, of the
order $10^{18}/{\rm cm}^2s$ and that the signal ($nn$ scattering) goes as the
square of the flux, while the background (neutron scattering against the
surroundings) is linear in the flux.
This work is still in progress.

\section{Conclusions and outlook}
During the last few decades the theoretical tools necessary for a precise
determination of the neutron-neutron scattering length has seen a remarkable
improvement.
First, with the modern computers it is possible to do a full Faddeev
calculation of the three-nucleon system with high-precision phenomenological
potentials.
Second, the EFTs provide a systematic and consistent framework for few-body
systems, with well-defined theoretical errors.
This has made it possible to put the three- and four-nucleon forces within the
same expansion scheme as the nucleon-nucleon force.
Also, the $\chi$PT scheme has established connections between various types of
reactions, through symmetries and LECs, which in turn makes it possible
to calculate seemingly intractable processes.
This makes it possible to constrain the calculations of
$\pi^-d\to nn\gamma$ by the LEC $\hat{d}$ extracted from tritium beta decay
(after incorporating it in the EFT calculation of the proton fusion process)
and thus help reduce the theoretical error in $a_{nn}$ extracted from the pion
capture process.
It is the hope of the author that the measurement of $\mu^-d\to nn\nu_\mu$
would provide an alternative way to constrain $\pi^-d\to nn\gamma$, now
directly from a two-body observable.

On the experimental side the $nd$ breakup experiments need to converge on a
consistent result.
The present situation with two conflicting $a_{nn}$ values is
highly unsatisfactory.
A solution might be in sight through the combined effort of both teams,
currently being pursued at TUNL, where the Bonn and TUNL experiment are being
set up together, using the same beam.

The $\pi^-d\to nn\gamma$, which so far has been considered the most promising
process, is no longer experimentally accessible, since all pion beam facilities
have shut down.
Its cousin $\gamma d\to nn\pi^+$ can be measured at TUNL
once the appropriate mirrors have been developed and produced.
An alternative would be using the tagged-photon experiment at MAXLab.

The most interesting prospect for $a_{nn}$ is the possibility, finally,
of a direct measurement at the pulsed reactor YAGUAR.

\section*{Acknowledgment}
I appreciate discussions with Christoph Hanhart, Calvin Howell,
Kuniharu Kubodera, Fred Myhrer, Daniel Phillips, and Henryk Wita{\l}a.
This work was supported by the U.S. National Science Foundation through the
the grants PHY-0457014 and PHY-0758114.

\end{document}